\documentclass[twocolumn]{aastex631}
\usepackage{graphicx}
\usepackage{float}
\usepackage{lineno}

\newcommand{\lsim }{{\lower0.8ex\hbox{$\buildrel <\over\sim$}}}
\newcommand{\gsim }{{\lower0.8ex\hbox{$\buildrel >\over\sim$}}}
\newcommand{\Msun}{\ifmmode {M_{\odot}}\else${M_{\odot}}$\fi}
\newcommand{\Lsun}{\ifmmode {L_{\odot}}\else${L_{\odot}}$\fi}
\newcommand{\Rsun}{\ifmmode {R_{\odot}}\else${R_{\odot}}$\fi}

\newcommand{\U}[3]{$#1_{-#3}^{+#2}$}

\graphicspath{{./}{images/}}

\shorttitle{X1850-087}
\shortauthors{Panurach et al.}

\begin{document}

\title{Tracking the Enigmatic Globular Cluster Ultracompact X-ray Binary X1850--087: Extreme Radio Variability in the Hard State}

\correspondingauthor{Teresa Panurach}
\email{panurach@msu.edu}

\author[0000-0001-8424-2848]{Teresa Panurach}
\affiliation{Center for Data Intensive and Time Domain Astronomy, Department of Physics and Astronomy, Michigan State University, East Lansing MI, USA}

\author[0000-0003-1814-8620]{Ryan Urquhart}
\affiliation{Center for Data Intensive and Time Domain Astronomy, Department of Physics and Astronomy, Michigan State University, East Lansing MI, USA}

\author[0000-0002-1468-9668]{Jay Strader}
\affiliation{Center for Data Intensive and Time Domain Astronomy, Department of Physics and Astronomy, Michigan State University, East Lansing MI, USA}

\author[0000-0002-8400-3705]{Laura Chomiuk}
\affiliation{Center for Data Intensive and Time Domain Astronomy, Department of Physics and Astronomy, Michigan State University, East Lansing MI, USA}

\author[0000-0003-2506-6041]{Arash Bahramian}
\affiliation{International Centre for Radio Astronomy Research Curtin University, GPO Box U1987, Perth, WA 6845, Australia}

\author[0000-0003-3944-6109]{Craig O. Heinke}
\affiliation{Department of Physics, University of Alberta, CCIS 4-181, Edmonton, AB T6G 2E1, Canada}

\author[0000-0003-0976-4755]{Thomas J. Maccarone}
\affiliation{Department of Physics \& Astronomy, Texas Tech University, Box 41051, Lubbock, TX 79409-1051, USA}

\author[0000-0003-2506-6041]{James C. A. Miller-Jones}
\affiliation{International Centre for Radio Astronomy Research Curtin University, GPO Box U1987, Perth, WA 6845, Australia}

\author[0000-0001-6682-916X]{Gregory R. Sivakoff}
\affiliation{Department of Physics, University of Alberta, CCIS 4-181, Edmonton, AB T6G 2E1, Canada}

\begin{abstract}
The conditions under which accreting neutron stars launch radio-emitting jets and/or outflows are still poorly understood. The ultracompact X-ray binary X1850--087, located in the globular cluster NGC 6712, is a persistent atoll-type X-ray source that has previously shown unusual radio continuum variability. Here we present the results of a pilot radio monitoring program of X1850--087 undertaken with the Karl G. Jansky Very Large Array, with simultaneous or quasi-simultaneous Swift/XRT data obtained at each epoch. The binary is clearly detected in the radio in two of the six new epochs. When combined with previous data, these results suggest that X1850--087 shows radio emission at a slightly elevated hard state X-ray luminosity of $L_X \gtrsim 2 \times 10^{36}$ erg s$^{-1}$, but no radio emission in its baseline hard state $L_X \sim 10^{36}$ erg s$^{-1}$. No clear X-ray spectral changes are associated with this factor of $\gtrsim 10$ radio variability. At all detected epochs X1850--087 has a flat-to-inverted radio spectral index, more consistent with the partially absorbed optically thick synchrotron of a compact jet rather than the evolving optically thick to thin emission associated with transient expanding synchrotron-emitting ejecta. If the radio emission in X1850--087 is indeed due to a compact jet, then it is plausibly being launched and quenched in the hard state on timescales as short as a few days. Future radio monitoring of X1850--087 could help elucidate the conditions under which compact jets are produced around hard state accreting neutron stars.
\end{abstract}

\section{Introduction} \label{sec:intro}
Due to their high stellar densities, globular clusters are known to host an abundant population of dynamically-formed low-mass X-ray binaries \citep{Clark75, Katz75, Ivanova07}. This is especially true for ultracompact X-ray binaries---a special class of low-mass X-ray binaries with $<1$ hr orbital periods that have hydrogen-poor donors (typically white dwarfs; \citealt{Nelson86}). Of the ten Galactic systems generally accepted as persistently accreting ultracompact X-ray binaries, four are in globular clusters {\citep{Sewardetal1976,Stella87,Heinke01,Zurek09}}. All of these cluster ultracompact binaries have very short orbital periods in the range $\sim 11$--21 min and neutron star primaries \citep{Heinke13}. 

Ultracompact X-ray binaries can form via standard common envelope evolutionary channels (see the review of \citealt{Nelemans10a}) or, in clusters, via direct collisions between neutron stars and red giants \citep{Ivanova05}. Neutron star ultracompact X-ray binaries are of special interest, as their accretion physics is expected to be relatively simple compared to the bulk of low-mass X-ray binaries. After formation, their evolution is primarily driven by gravitational wave emission rather than complex stellar physics \citep{Suvorov21, Chen21}. Ultracompact binaries with periods $\lesssim 25$ min are expected to have high enough accretion rates to keep the disk fully ionized at essentially all times and hence in a state of persistently high luminosity, $L_X \sim 10^{36}$ erg s$^{-1}$ (e.g., \citealt{Deloye2003, Lasota2008}). All of the globular cluster-hosted ultracompact X-ray binaries meet this orbital period criterion and show satisfying agreement between the mean predicted and observed mass transfer rates \citep{Heinke13}. However, they also show both short (days) and long-term (several months) variations in their X-ray luminosities that are not well understood \citep{Simon03,intZand07,Maccarone10}. Radio emission from accreting neutron stars, including a few ultracompact X-ray binaries, has also been shown to vary on similar time scales \citep{diaztrigo2017,Gusinskaia20a,Russell21}.

X1850–087 in NGC 6712 is one of the four persistent ultracompact neutron star X-ray binaries in globular clusters \citep{Sewardetal1976}, with a suspected 20.6-min orbital period \citep{Homer96} and distance of 8.0 kpc \citep{Tremou18}. As part of the MAVERIC radio continuum survey of Galactic globular clusters \citep{Tremou18,Shishkovsky20,Tudor22}, X1850–087 was observed over seven epochs, each of 1--2 hour duration, in April and May 2014 \citep{Panurach21}. In these data an unusual degree of radio variability was observed, with a mixture of stringent upper limits ($< 12-15\ \mu$Jy at 5.0 and 7.4 GHz) and relatively bright detections ($\sim 130-200\ \mu$Jy at the same frequencies). This variability was difficult to interpret owing to the lack of simultaneous X-ray data for most of the radio epochs, other than MAXI and Swift/BAT data that require coarse time binning to detect X1850–087. There was some evidence that X1850–087 brightened in the X-rays (by a factor of $\sim 2$) and became somewhat softer during the timespan of the radio observations, but this does not fully explain the radio variability on timescales of a few days \citep{Panurach21}.

\citet{Panurach21} also analyzed archival VLA observations of X1850–087 from 1989 to 1998, finding radio continuum detections comparably luminous to those in 2014 in two of the three archival epochs. Again there are no simultaneous X-ray data for these older observations. But given that the long-term X-ray light curve of the source from RXTE shows that it is only in an obvious bright state ($L_X \gtrsim 2-3 \times 10^{36}$ erg s$^{-1}$) for
$6-15\%$ of the time, it would be unlikely to have taken several additional independent radio observations during this uncommon X-ray bright/flaring state. Instead, it suggested that unexpectedly luminous radio continuum emission might be characteristic of X1850–087 at its typical X-ray luminosity of 
$L_X \sim 10^{36}$ erg s$^{-1}$ \citep{Panurach21}.

Since it was clear that a lack of simultaneous or quasi-simultanous X-ray data marred one's ability to interpret the existing radio observations, we undertook a monitoring campaign of X1850-087 that included both radio and X-ray observations, lasting from Feb to June 2022. This paper describes and interprets the results of this campaign.

\section{Data} \label{sec:2}
\subsection{Radio Observations}
We obtained new Very Large Array (VLA) observations of X1850-087 over six well-spaced epochs from 2022 February to July as part of NRAO program VLA/22A-111. All observations were made in the most extended A configuration using C band receivers (4.0--8.0 GHz), split into two 2.0 GHz wide subbands centered at 5.0 and 7.0 GHz. We used 3C286 or 3C48 as the flux calibrator and J1832--1035 as the complex gain calibrator for each 30-min observing block.

The data were flagged and calibrated using the standard reduction pipeline in {\sc CASA} ver.5.6.2 \citep{CASA2022} and imaged using {\sc CASA}'s \verb|tclean| feature using the Briggs weighting scheme of robustness 1. 

Radio flux density measurements were made using {\sc CASA}'s \verb|imfit| task by fitting a point source model matching the synthesized beam. If the source was not detected at a given epoch and frequency, we report upper limits at the $3\sigma$ level, where $\sigma$ is defined as the local rms noise. Details of the radio observations and the measured flux densities or upper limits are listed in Table \ref{tab:x1850}.

\subsection{X-ray Observations}

In conjunction with our radio monitoring of X1850-087, we also obtained quasi-simultaneous X-ray data with \emph{Swift}/XRT, with the interval between the radio and X-ray observations ranging from $\sim 5$ hr to $\sim 2$ d (see Table \ref{tab:x1850}). Owing to the high X-ray flux of X1850-087, all \emph{Swift}/XRT observations were made in windowed timing mode. All observations were analyzed using software within the {\sc ftools}\footnote{http://heasarc.gsfc.nasa.gov/ftools} package \citep{Blackburn95}. The XRT data were reprocessed using the \verb|xrtpipeline| task. Following standard procedures, light curves and spectra were extracted with \verb|xselect|. Due to a 5 ksec gap in the 2022 May 20 observation, we split it into two different datasets and analyzed them independently. All spectra were grouped to have at least 15 counts per bin so that Gaussian statistics could be used. Spectral fitting was performed using XSPEC version 12.11.1 \citep{Arnaud96}.


\begin{figure*}[!t]
  \centering
 \includegraphics[width=1.0\textwidth]{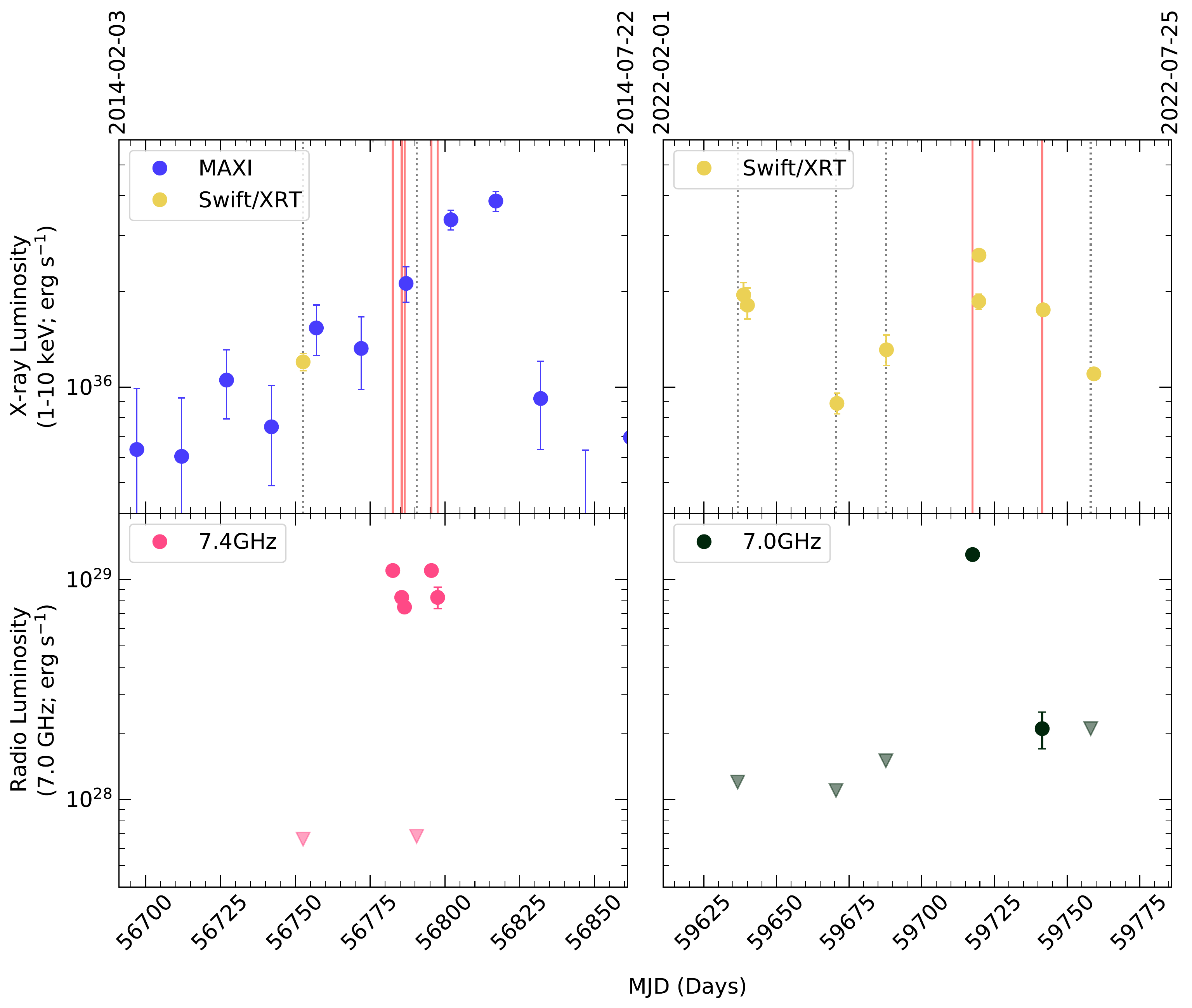}
  \caption{2014 (left panel) and 2022 (right panel) radio and X-ray lightcurves for X1850--087 (see also \citealt{Panurach21}). The top row shows X-ray lightcurves from MAXI (blue) and Swift/\emph{XRT} (yellow), all plotted as 1--10 keV luminosities. The bottom row shows 7.4 (left in pink; from 2014) and 7.0 GHz (right in black; from 2022) VLA radio luminosity lightcurves. Filled circles represent detections while upside down triangles are $3\sigma$ upper limits. Solid salmon vertical lines in the top row represent radio detections, while gray dotted lines represent non-detections.}
 \label{fig:f1}
  \end{figure*}

\begin{figure*}
    \centering
    \includegraphics[width=0.96\textwidth]{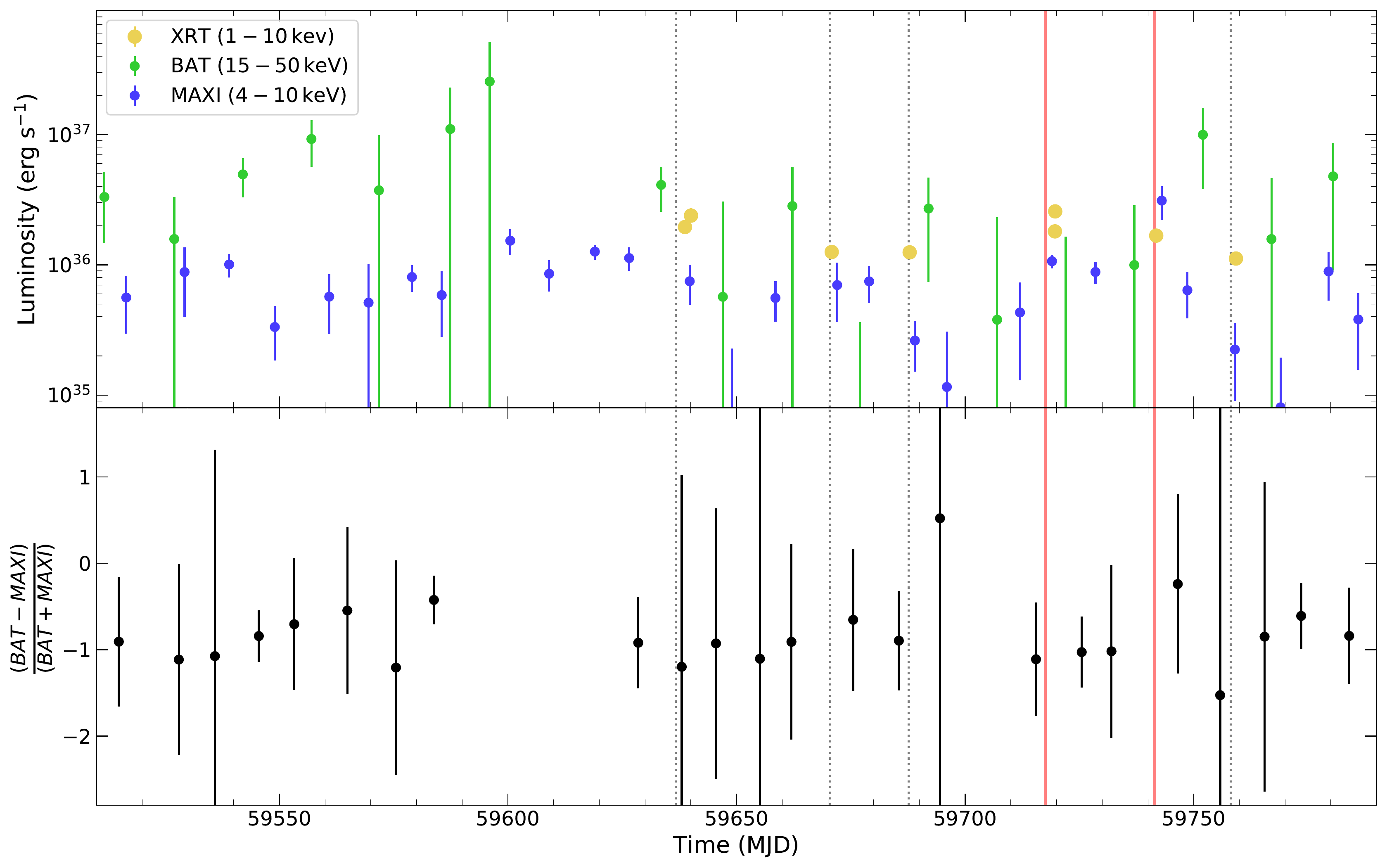}
    \caption{Top: X-ray light curves for MAXI (4-10 keV; blue) and \textit{Swift}/BAT (15-50 keV; green) for times surrounding our 2022 VLA data. MAXI data points are binned to 10 days while BAT data are binned to 15 days. Also plotted are the Swift/XRT measurements (1--10 keV; yellow). Dashed vertical grey lines indicate non-detections in our simultaneous VLA observations; the solid vertical red line represents radio detections. Bottom: Hardness ratio light curve for data in top panel.}
    \label{fig:hardness}
\end{figure*}

\section{Results} \label{sec:3}

\subsection{Radio}
In the first three observations (2022 Feburary 26, 2022 April 01, and 2022 April 18), X1850-087 was undetected at both 5.0 and 7.0 GHz, with typical $3\sigma$ upper limits in the range $\lesssim 20$--30 $\mu$Jy. In the fourth observation, obtained on 2022 May 18, the source was well-detected, with 5.0 and 7.0 GHz flux densities of $212\pm9$ and $226\pm9\ \mu$Jy, respectively. For a power-law $S_{\nu} \propto \nu^{\alpha}$ with flux density $S_{\nu}$ at frequency $\nu$, these measurements give a flat to slightly-inverted spectral index of $\alpha = 0.19 \pm 0.17$. 

In the next epoch, on 2022 June 11 (24 days later), the source was still detected at 7.0 GHz ($39\pm9\ \mu$Jy), but was about a factor of 6 fainter than on May 18. It was undetected at 5.0 GHz ($3\sigma$ upper limit of $< 32\ \mu$Jy). Finally, in the last epoch on 2022 June 28, it was not detected at either frequency ($< 40.5$ and $< 39.6\ \mu$Jy at 5.0 and 7.0 GHz).

For the four epochs without a detection at either frequency, we averaged the subbands together to lower the noise, but still did not detect the source at any of these epochs.



\subsection{X-ray}

The source is well-detected with \emph{Swift}/XRT in all of the X-ray epochs. As a starting model, we initially fit each of the observations with a single-component absorbed blackbody disk model with both local and Galactic absorption (\textit{TBabs}$\times$\textit{TBabs}$\times$\textit{diskbb}). This provided a poor fit across all epochs.

The fits are greatly improved by instead using an absorbed power-law model (\textit{TBabs}$\times$\textit{TBabs}$\times$\textit{powerlaw}). While the power-law models provide suitable fits, some epochs show minor soft residuals, which we fit with an additional low temperature \textit{diskbb} component. A summary of our spectral results is given in Table \ref{tab:xray_spectral}. The two-component model (\textit{diskbb+powerlaw}) is only used in observations where the addition of the \textit{diskbb} is  a statistically significant improvment ($\geq 99$\% confidence level) relative to the single-component powerlaw model. The power-law component dominates the thermal component in all cases, implying that X1850--087 was always in a hard state during these observations.

We find small but statistically significant variations in the $N_H$ intrinsic to the source in the X-ray spectral fits. Given that previous high signal-to-noise X-ray observations have also found evidence for $N_H$ variations in X1850--087, where they are mooted to arise from a variable disk wind \citep{Sidoli2005}, we think these inferred $N_H$ variations are likely to be real, though as expected the precise value depends on the spectral model fit.

\section{Discussion} \label{sec:4}

\subsection{Emission at Baseline in X1850-087}

One of the outstanding questions posed by existing radio observations of X1850-087 is whether the source shows radio emission during its normal persistent hard state at $L_X \sim 10^{36}$ erg s$^{-1}$ (1--10 keV). In April 2014, it was undetected in deep VLA radio continuum observations at this luminosity.


Our new VLA and quasi-simultaneous \emph{Swift} data provide suggestive (but not yet conclusive) evidence that X1850--087 does not have a persistent compact radio jet at its baseline X-ray luminosity. In three of the six new epochs, the source is around $L_X \sim 10^{36}$ erg s$^{-1}$, and there is no detectable radio emission, as was the case for the VLA data taken at a similar $L_X$ in April 2014 (radio upper limit $< 12.9 \mu$Jy at 5.0 GHz). An archival VLA non-detection from 1998 provides only a weak radio upper limit ($< 135 \mu$Jy at 4.9 GHz), while quasi-simultaneous X-ray data from the Rossi X-ray Timing Explorer show the X-ray luminosity appears close to the mean baseline level at this epoch. 

\citet{Panurach21} also presented results from two even earlier archival VLA observations (from 1989 and 1991), in which X1850--087 is clearly detected at 4.9 GHz in each ($156 \pm 24$ $\mu$Jy and $134 \pm 23$ $\mu$Jy, respectively). Unfortunately,  there are no X-ray observations that are even quasi-simultaneous with these archival radio detections, so it is not possible to tell whether they occurred during periods of elevated X-ray luminosity.




While additional simultaneous X-ray and radio observations would be useful, at present there is no countervailing evidence to the statement that X1850--087 does not have a persistent compact radio jet/outflow in its baseline state.


\subsection{Origin of Radio Emission: A Variable Jet?} \label{sec:4.2}

The May 2014 VLA observations of X1850--087 showed a clear detection of luminous radio emission in five different epochs over a 15-day timespan, mixed in with a single deep radio upper limit 4 days after (and 5 days before) strong detections at a flux density $\sim 20\times$ higher than the $3\sigma$ upper limit (Figure \ref{fig:f1}; bottom left). Unfortunately, no targeted simultaneous X-ray data are available for these epochs, but time-binned MAXI data show evidence for an increase in the X-ray luminosity during this timespan, peaking at $\sim 4 \times 10^{36}$ erg s$^{-1}$ two weeks after the last radio observation (Figure \ref{fig:f1}; top left). Hence it seems generally plausible to suggest that X1850--087 can emit luminous radio emission at $L_X \gtrsim 2\times10^{36}$ erg s$^{-1}$, with the important caveat that the radio non-detection in the midst of the 2014 detections must also be explained.
 
The 2022 data show a similar variability, though with some unanswered questions. X1850--087 is indeed strongly detected in the radio quasi-simultaneously with the most luminous X-ray emission observed over this timespan, at $L_X = 2.6 \times10^{36}$ erg s$^{-1}$ in the second \emph{Swift}/XRT observation on 2022 May 20 (Figure \ref{fig:f1}; right). It is still detected in the radio three weeks after this epoch, at a still-elevated X-ray luminosity of $\sim 1.6 \times10^{36}$ erg s$^{-1}$, but at the lowest radio flux density for any detection ever made of this source: $39\pm9 \mu$Jy at 7.0 GHz (it was not detected at 5.0 GHz). Given the evidence from 2014 for strong evolution of the radio flux density in X1850--087 over timescales as short as a few days, it seems plausible that the second, fainter detection is due to a radio jet or outflow that is causally unrelated to the bright radio emission seen three weeks previously. However, this cannot be definitively determined with these data.

In all of the 2014 and 2022 radio-detected epochs the radio spectral index is flat to inverted, consistent with partially self-absorbed optically-thick synchrotron emission as expected for a compact jet. Hence one interpretation superficially consistent with the data is that a compact jet is present in X1850--087 in the hard state only at $L_X \gtrsim 2 \times 10^{36}$ erg s$^{-1}$, which corresponds to an accretion rate of $\sim 3 \times 10^{-10} M_{\odot}$ yr$^{-1}$ (scaling from the results of \citealt{Heinke13}). Launching of the jet at around $L_X \gtrsim 2\times10^{36}$ erg s$^{-1}$ could also self-consistently explain the finding in \citet{Panurach21} that X1850--087 is detected at comparable radio luminosities in archival VLA data from 1989 and 1991. These observations were more difficult to explain in a model where the radio emission was only observable during rarer distinct periods when $L_X > 3\times10^{36}$ erg s$^{-1}$ rather than more common fluctuations to $L_X \sim 2\times10^{36}$ erg s$^{-1}$  in the normal persistent hard state.

In the jet interpretation, at lower accretion rates the jet is either not launched or is much less luminous than would be expected if the system followed any of the $L_R$--$L_X$ relations proposed for other accreting neutron stars in the hard state. For example, even assuming a relatively steep $L_R \propto L_X^{1.4}$ \citep{Migliari06}, only a factor of 2.5--3.5 radio variability would be expected for X1850--087 in our data, rather than the factor of $\gtrsim 20$ observed. Hence, in this interpretation a quenching of the jet at \emph{low luminosity} would be required.

An alternative explanation is that X1850--087 is emitting discrete transient synchrotron blobs, perhaps connected with minor variations in its spectral state or luminosity. While this hypothesis is difficult to fully rule out, in broad strokes it is not consistent with our data. On 2022 May 18, when the source is brightest in the radio with a 7.0 GHz flux density of $226\pm9 \mu$Jy, the spectral index is $\alpha = +0.19\pm0.17$, suggesting partially optically thick synchrotron emission. 24 days later, even though the 7.0 GHz flux density has fallen by a factor of $\sim 6$ to $39\pm9 \mu$Jy, the spectral index is still inverted with $\alpha \gtrsim 0$, as it is not detected at all in the 5.0 GHz image ($3\sigma$ upper limit $<31.5 \mu$Jy). This behavior is contrary to that expected for an expanding synchrotron blob, which
would be expected to become optically thin and hence brighter at lower frequencies as it fades.

The 2014 observations are also more consistent with a compact jet than synchrotron ejecta: in 5 detected epochs, taken over 16 days, the spectral index is always flat to inverted and never optically thin. In addition, in these epochs the 7.0 GHz flux density sits in a narrow range of $\sim 150-200 \mu$Jy, rather than the broader range that would be expected for expanding blobs.

\subsection{Caveats and Comparisons}

We have suggested a scenario that appears to explain the bulk of the observations of X1850--087: a compact jet that produces radio emission is sometimes present in the persistent hard state, but only above a threshold X-ray luminosity of $L_X \gtrsim 2\times10^{36}$ erg s$^{-1}$. 

The first caveat to this scenario is that it requires jet launching and quenching to be able to occur on timescales as short as those observed in our 2014 VLA data, 4--5 days. Jet quenching on such timescales has been observed during hard-to-soft state transitions of neutron star low-mass X-ray binaries (e.g., \citealt{MillerJones10}), but quenching without a clear state change as we propose for X1850--087 would likely require a different physical mechanism. 

Another important uncertainty is whether an accretion rate onto the neutron star in that produces $L_X \gtrsim 2\times10^{36}$ erg s$^{-1}$ is a sufficient condition to produce detectable radio emission from X1850--087 or whether other, perhaps stochastic factors, are also at play. The first 2022 X-ray observation in our dataset (2022 Feb 28) has an inferred $L_X = (2.0\pm0.2) \times 10^{36}$ erg s$^{-1}$, with a similar luminosity in X-ray data obtained 1.3 d later, but there is no radio continuum emission detected in the quasi-simultaneous VLA observations. The spectral state of X1850--087 is not too dissimilar between these epochs and those on 2022 May 20 when the system is radio-luminous: formally a thermal disk improves the X-ray spectral fit in the earlier observations and not in the latter ones, but this carries only a small fraction of the X-ray luminosity in the Swift/XRT band, and does not represent strong evidence for a spectral state change. The MAXI and Swift/BAT data also show no clear evidence for spectral variations, though they are coarsely binned in time (Figure \ref{fig:hardness}). One possibility is simply that the threshold X-ray luminosity is slightly above $2\times10^{36}$ erg s$^{-1}$, such that it is met on 2022 May 20 and not on 2022 Feb 28. Nothing in the 2014 data contradicts this idea---all the radio-bright epochs appear to be associated with $L_X > 2\times10^{36}$ erg s$^{-1}$---
but this 2014 finding relies on coarsely time-binned MAXI data.

Comparing X1850--087 to other accreting neutron star binaries, 
we note that similar radio variability and threshold emission behaviors have been found in some other systems, though potentially in different contexts. In Be/X-ray binary system Swift J0243.6+6124, radio emission was only detected above $L_X > 2\times10^{36}$ erg s$^{-1}$ \citep{vandenEijnden19} assuming a distance of 5.0 kpc \citep{vandenEijnden18}. Since the neutron star in this binary is inferred to have an extremely high magnetic field of $B\sim 10^{13}$ G \citep{Wilson-Hodge2018}, \citet{vandenEijnden18} suggested the threshold luminosity was due to a magnetic centrifugal barrier suppressing radio emission at lower accretion rates.
While the magnetic field strength of the neutron star in X1850--087 is unknown, an old, at least partially recycled globular cluster neutron star is much more likely to have a lower magnetic field, $B \lesssim 10^{10}$ G (e.g., \citealt{Revnivtsev15}), and hence a much lower threshold luminosity. This casts doubt on the relevance of this mechanism for X1850--087.

A few accreting millisecond X-ray pulsars have also shown variability and/or potentially threshold behavior in the hard state. For example, Aql X-1 has a significant number of 
radio and X-ray studies \citep{Migliari06, Tudose09,MillerJones10,Gusinskaia20a} in which it shows factor of $\sim 4$ radio variability. Aql X-1 has no radio detections in the hard state at $L_X \lesssim 10^{36}$ erg s$^{-1}$, with a range of upper limits of differing depths, along with at mixture of radio detections and upper limits 
in the range $L_X \sim 10^{36}$ erg s$^{-1}$ to $3 \times 10^{36}$ erg s$^{-1}$ (assuming a distance of 4.5 kpc; \citealt{Gungor17}). These data could conceivably be consistent with a luminosity threshold for radio emission, but instead also with a steep radio--X-ray relation and radio variability; additional deep simultaneous X-ray and radio data are needed. 
Two other accreting millisecond X-ray pulsars, SAX J1808.4--3658 and IGR J00291+5934, have been detected in the radio at multiple epochs with $L_X$ between $10^{34}$ and $10^{36}$ erg s$^{-1}$ assuming a distance of 3.5 kpc and 4.2 kpc respectively; \citealt{Tudor17}). Hence, these binaries do not evince a clear threshold for radio emission, though they do show radio variability of at least a factor of a few \citep{Tudor17}.

As potentially a more direct comparison, the ultra-compact X-ray binary 4U 0614+091 has a similar baseline X-ray luminosity and spectral shape to X1850--087, but appears to have a luminous radio jet at this luminosity, albeit with a limited number of clear radio detections \citep{Migliari10}.

\section{Summary and Future Work}

In this paper we present new radio and X-ray monitoring observations taken from February to July 2022 of the ultracompact neutron star X-ray binary X1850--087 in NGC 6712 to better understand the radio variability previously observed in 2014. In these new data, we observe both extreme radio variability and a flat to slightly inverted radio spectral index in the detected epochs, consistent with the 2014 radio data. We now also have quasi-simultaneous X-ray data for all epochs to aid in interpreting these new radio observations, unlike the more fragmented X-ray data previously available. We find that detectable radio emission from X1850--087 in the hard state seems to only occur at $L_X \gtrsim 2 \times 10^{36}$ erg s$^{-1}$.

Given that the properties of the radio emission at all detected epochs support the presence of a compact jet, a plausible interpretation of the radio/X-ray behavior of X1850--087 is that the source has a variable compact jet in the hard state that is only launched above a specific threshold X-ray luminosity. However, due to the limited X-ray coverage of previous datasets, we cannot rule out a scenario where there is no (or a much lower) threshold for jet launching, combined with an unusual level of radio variability in a restricted range of X-ray luminosities. A much less likely variation on this scenario is that the radio variability closely tracks the X-ray variability; this scenario is disfavored as it would require an implausibly steep radio/X-ray correlation. Scenarios where the radio emission is due to outflows of synchrotron blobs rather than a compact jet are strongly inconsistent with the behavior of the radio spectral indices and hence also disfavored.

It is clear that additional time-intensive quasi-simultaneous radio/X-ray observations of X1850--087 are needed to better characterize the range of its behaviors and test the scenarios discussed here. In addition to X1850--087, we highlight the persistent source 4U 0614+091 and the frequently outbursting source Aql X-1 as other neutron star binaries for which improved radio characterization in the hard state could be especially revealing.

\section{Acknowledgements} \label{sec:5}

We acknowledge the helpful comments of an anonymous referee.

JS acknowledges support from NASA grant 80NSSC21K0628 and the Packard Foundation.
TP is supported by the National Science Foundation Graduate Research Fellowship under Grant No.~DGE-2039655.

The National Radio Astronomy Observatory is a facility of the National Science Foundation operated under cooperative agreement by Associated Universities, Inc. We acknowledge the use of public data from the Swift data archive. This research has made use of MAXI data provided by RIKEN, JAXA and the MAXI team.\\

\noindent \textit{Facilities}: VLA, \emph{Swift}/XRT, MAXI

\noindent \textit{Software}: Astropy \citep{Robitaille13}, \textsc{CASA} \citep{CASA2022}, \textsc{FTOOLS} \citep{Blackburn95}, \textsc{Matplotlib} \citep{Hunter07}, \textsc{NumPy} \citep{harris20}, \textsc{pandas} \citep{Mckinney10},   \textsc{Xspec} \citep{Arnaud96}

\begin{deluxetable*}{lrrrrrrr} 
\label{tab:x1850}
\tabletypesize{\footnotesize}
\tablewidth{0pt} 
\tablecaption{2022 Radio and X-ray Observations} 
\tablehead{Radio Observation Date & 5.0GHz Flux Density & 7.0GHz Flux Density & X-Ray Date &  X-ray Flux (1 - 10 keV)\\
(Midpoint UT) &($\mu$Jy) & ($\mu$Jy) & (Midpoint UT) &($10^{-10}$ ergs s$^{-1}$ cm$^{-2}$) }
\startdata
2022 Feb 26 (15:49:37)  & $< 30.0$ & $< 22.5$ & 2022 Feb 28 (17:13:08.09)
 & $2.5\pm0.2$\\
&  &  &  2022 Mar 2 (00:48:14.12) & \U{2.4}{0.3}{0.2} \\
2022 Apr 01  (12:13:31) & $< 30.3$ & $< 21.3$ & 2022 Apr 01 (19:32:46.54) & $1.16\pm0.09$\\
2022 Apr 18 (15:48:18) & $< 32.7$ & $< 29.7$ & 2022 Apr 18 (20:40:28.68)& \U{1.71}{0.19}{0.18} \\
2022 May 18  (12:03:25) & $212 \pm 8.9$ & $226.2 \pm 8.9$ & 2022 May 20 (15:14:00.12) & $2.43\pm0.13$ \\
 &  & & 2022 May 20 (16:43:32.20) & $3.39\pm0.10$ \\
2022 Jun 11  (10:27:36) & $< 31.5$ & $39.1 \pm 9.0$ & 2022 Jun 11 (18:58:43.63)  & \U{2.28}{0.09}{0.08} \\
2022 Jun 28 (03:20:37) & $< 40.5$ & $< 39.6$ & 2022 Jun 29 (05:32:34.28) & $1.43\pm0.06$  \\
 \enddata
\end{deluxetable*}


\begin{deluxetable*}{llrrrrrrrr}
    \caption{X-ray spectral properties of X1850--087}
    \tablecolumns{10}
    \tablewidth{0pt}
    \rotate
    \renewcommand{\arraystretch}{1.4}
    \tablehead{
    \hline
    \multicolumn{1}{c}{Comp} & \multicolumn{1}{c}{Parameter} & \multicolumn{8}{c}{Observation ID} \\  
      & & \multicolumn{8}{c}{Date} \\  
    & & \multicolumn{1}{c}{15041001} & \multicolumn{1}{c}{15041002} & \multicolumn{1}{c}{15041003} & \multicolumn{1}{c}{15041004} & \multicolumn{1}{c}{15179001a} & \multicolumn{1}{c}{15179001b} & \multicolumn{1}{c}{15213001} & \multicolumn{1}{c}{15213001}\\
     & & \multicolumn{1}{c}{2022 Feb 28} & \multicolumn{1}{c}{2022 Mar 2} & \multicolumn{1}{c}{2022 Apr 1} & \multicolumn{1}{c}{2022 Apr 18} & \multicolumn{1}{c}{2022 May 20} & \multicolumn{1}{r}{2022 May 20} & \multicolumn{1}{r}{2022 Jun 11} & \multicolumn{1}{r}{2022 Jun 29}
    }
    \startdata
    \textit{TBabs} & $N_{\rm H,\,LoS} \,(10^{22} {\rm cm}^{-2})$ &  [0.25] & [0.25] & [0.25] & [0.25] & [0.25] & [0.25] & [0.25] & [0.25] \\\hline
    \textit{TBabs} & $N_{\rm H,\,int} \,(10^{22} {\rm cm}^{-2})$ &  $<0.02$ & $0.09\pm0.03$ & \U{0.19}{0.09}{0.08} & $0.08\pm0.04$ & $0.07\pm0.04$ & $0.05\pm0.02$ & $<0.02$ & $<0.01$ \\
    \textit{powerlaw} & $\Gamma$ & \U{2.31}{0.08}{0.04} & $2.78\pm0.13$ & $2.2\pm0.2$ & $2.2\pm0.1$ & \U{2.66}{0.15}{0.14} & $2.55\pm0.08$ & $2.53\pm0.05$ & $2.35\pm0.05$ \\
     & normalization & \U{0.078}{0.006}{0.002} & \U{0.10}{0.04}{0.02} & \U{0.040}{0.009}{0.007} & $0.08\pm0.02$ & \U{0.128}{0.017}{0.015} & \U{0.163}{0.011}{0.010} & $0.074\pm0.010$ & $0.056\pm0.002$\\
    & $f^{tot}_{1-10\,\rm{keV}}$ ($10^{-10}$ erg s$^{-1}$ cm$^{-2}$) & $2.05\pm0.07$ & $2.09\pm0.10$ & $1.16\pm0.09$ & $1.40\pm0.06$ & $2.43\pm0.13$ & $3.39\pm0.10$ & $2.05\pm0.04$ & $1.43\pm0.06$ \\
    & $L^{tot}_{1-10\,\rm{keV}}$ ($10^{36}$ erg s$^{-1}$) & $1.57\pm0.05$ & $1.60\pm0.07$ & $0.89\pm0.07$ & $1.08\pm0.05$ & $1.86\pm0.10$ & $2.60\pm0.08$ & $1.57\pm0.03$  & $1.10\pm0.05$ \\
    & $\chi^2_{\rm{red}}$ (degrees of freedom) & 1.01(231) & 1.08(156) & 0.89(75) & 1.13(168) & 1.26(108) & 1.17(238) & 1.64(341) & 0.94(195) \\\hline
    \textit{TBabs} & $N_{\rm H,\,int} \,(10^{22} {\rm cm}^{-2})$ &  \U{0.37}{0.13}{0.15} & \U{0.22}{0.18}{0.14} & & \U{0.48}{0.18}{0.19} &  & & \U{0.12}{0.06}{0.05} & \\
    \textit{powerlaw} & $\Gamma$ & $2.50\pm0.15$ & $2.6\pm0.3$ &  & $2.55\pm0.19$ & & & $2.28\pm0.11$ & \\
     & normalization & $0.10\pm0.02$ & \U{0.10}{0.04}{0.02} &  & $0.08\pm0.02$ & & & $0.074\pm0.010$ & \\
    \textit{diskbb} & $kT$\,(keV) & \U{0.120}{0.022}{0.014} & \U{0.16}{0.09}{0.06} &  & \U{0.091}{0.014}{0.010} &  &  & \U{0.18}{0.03}{0.02} &  \\
     & normalization ($10^4$) & \U{150}{760}{130} & \U{7.0}{148}{6.9} &  & \U{720}{4860}{650} &  &  & \U{2.8}{6.8}{2.0} &  \\
    & $f^{powerlaw}_{1-10\,\rm{keV}}$ ($10^{-10}$ erg s$^{-1}$ cm$^{-2}$) & $2.29\pm0.15$ & \U{2.1}{0.2}{0.3} & & $1.64\pm0.14$ & && \U{2.01}{0.09}{0.08} &\\
    & $f^{diskbb}_{1-10\,\rm{keV}}$ ($10^{-10}$ erg s$^{-1}$ cm$^{-2}$) & \U{0.25}{0.11}{0.09} & \U{0.30}{0.11}{0.15} & & \U{0.06}{0.07}{0.04} & && \U{0.27}{0.05}{0.04} &\\
     & $f^{tot}_{1-10\,\rm{keV}}$ ($10^{-10}$ erg s$^{-1}$ cm$^{-2}$) & $2.5\pm0.2$ & \U{2.4}{0.3}{0.2} &  & \U{1.71}{0.19}{0.18} &  & & \U{2.28}{0.09}{0.08} &  \\
     & $L^{tot}_{1-10\,\rm{keV}}$ ($10^{36}$ erg s$^{-1}$) & $2.0\pm0.2$ & $1.8\pm0.2$ &  & \U{1.31}{0.15}{0.14} & && \U{1.75}{0.07}{0.06}  & \\
     & $\chi^2_{\rm{red}}$ (degrees of freedom) & 0.90(229) & 0.94(154) &  & 1.06(166) &  & & 1.23(339) & \\\hline
     \enddata
     \tablecomments{The Galactic line-of-sight absorption ($N_{\rm H,\,LoS}$; values in square brackets) is fixed. All observations are initially fit with an absorbed powerlaw model. In some observations, an additional \textit{diskbb} component significantly (confidence level $>99\%$) improves the fit. Reported fluxes ($f_{1-10\,\rm{keV}}$) and luminosities ($L_{1-10\,\rm{keV}}$) are unabsorbed.}
    \label{tab:xray_spectral}
\end{deluxetable*}

\bibliography{references.bib}{}
\bibliographystyle{aasjournal}

\end{document}